\catcode`\@=11 
\def\lsim{\mathrel{\mathpalette\@versim<}}
\def\gsim{\mathrel{\mathpalette\@versim>}}
\def\@versim#1#2{\vcenter{\offinterlineskip
        \ialign{$\m@th#1\hfil##\hfil$\crcr#2\crcr\sim\crcr } }}

\documentstyle[preprint,aps]{revtex}
\begin{document}

\newcommand{\ee}{\mbox{$e^+e^-$}}
\newcommand{\phinaught}{\mbox{$\langle\Phi\rangle_0$}}
\newcommand{\mt}{m_t}
\newcommand{\mh}{m_H}
\newcommand{\mgev}{GeV/c$^2$}
\newcommand{\mevocsq}{MeV/c$^2$}
\newcommand{\mum}{$\mu$m}
\newcommand{\roots}{\sqrt{s}}
\newcommand{\alr}{A_{LR}}
\newcommand{\alro}{A_{LR}^0}
\newcommand{\swein}{\sin^2\theta_W^{\rm eff}}
\newcommand{\pole}{{\cal P}_e}
\newcommand{\polp}{{\cal P}_p}
\newcommand{\pollin}{{\cal P}_e^{lin}}
\newcommand{\pollum}{\bar{\cal P}_e}
\newcommand{\polg}{{\cal P}_\gamma}
\newcommand{\poll}{{\cal P}}
\newcommand{\apolel}{\langle{\cal P}_e\rangle}
\newcommand{\ameas}{A_m}
\newcommand{\lum}{{\cal L}}
\newcommand{\alum}{A_{\cal L}}
\newcommand{\aeff}{A_\varepsilon}
\newcommand{\apol}{A_{\cal P}}
\newcommand{\apower}{{\cal A}}
\newcommand{\aengy}{A_E}
\newcommand{\aback}{A_b}
\newcommand{\etal}{{\it et al.}}
\newcommand{\zo}{Z}
\newcommand{\rar}{\rightarrow}

\draft

\preprint{\vbox{\hsize=120pt\noindent SLAC--PUB--8401 \\
April 2000 }}

\title{A High-Precision Measurement of the Left-Right $\zo$
Boson Cross-Section
Asymmetry$^\dagger$}
\author{The SLD Collaboration$^*$}
\address{Stanford Linear Accelerator Center\\
         Stanford University, Stanford, California, 94309\\}
\maketitle

\begin{abstract}
We present a measurement of
the left-right cross-section asymmetry ($\alr$)
for $\zo$ boson production
by \ee\ collisions.  The measurement 
includes the final data taken
with the SLD detector at the SLAC Linear Collider (SLC)
during the period 1996-1998.
Using a sample of 383,487 $\zo$ decays collected during the
1996-1998 runs  we measure the pole-value
of the asymmetry, $\alro$, to be
0.15056$\pm$0.00239 which is equivalent
to an effective weak mixing angle of
$\swein=0.23107\pm0.00030$.
Our result for the complete 1992-1998 dataset 
comprising 537 thousand $\zo$ decays is 
$\swein=0.23097\pm0.00027$.

\end{abstract}
\vskip 0.3in
\begin{center}
{\rm Submitted to {\em Physical Review Letters}}
\end{center}
\vskip 1.3in
\vbox{\footnotesize\renewcommand{\baselinestretch}{1}\noindent
 $^\dagger$This work was supported in part by Department of Energy
  contract DE-AC03-76SF00515}
\pagebreak
\begin{center}
\bigskip
%
%
%
\def\iAOMORI{$^{(1)}$}
\def\iBRI{$^{(2)}$}
\def\iBRUN{$^{(3)}$}
\def\iBU{$^{(4)}$}
\def\iCOLO{$^{(5)}$}
\def\iCSU{$^{(6)}$}
\def\iFERR{$^{(7)}$}
\def\iFRAS{$^{(8)}$}
\def\iJHU{$^{(9)}$}
\def\iLBL{$^{(10}$}
\def\iMASS{$^{(11)}$}
\def\iMISSI{$^{(12)}$}
\def\iMIT{$^{(13)}$}
\def\iMOSCOW{$^{(14)}$}
\def\iNAGO{$^{(15)}$}
\def\iOREG{$^{(16)}$}
\def\iOXF{$^{(17)}$}
\def\iPERU{$^{(18)}$}
\def\iRAL{$^{(19)}$}
\def\iRUTG{$^{(20)}$}
\def\iSLAC{$^{(21)}$}
\def\iSOONG{$^{(22)}$}
\def\iTENN{$^{(23)}$}
\def\iTOHO{$^{(24)}$}
\def\iUCSB{$^{(25)}$}
\def\iUCSC{$^{(26)}$}
\def\iVAND{$^{(27)}$}
\def\iWASH{$^{(28)}$}
\def\iWISC{$^{(29)}$}
\def\iYALE{$^{(30)}$}

\baselineskip=.75\baselineskip 

\mbox{Kenji  Abe\unskip,\iNAGO}
\mbox{Koya Abe\unskip,\iTOHO}
\mbox{T. Abe\unskip,\iSLAC}
\mbox{I. Adam\unskip,\iSLAC}
\mbox{H. Akimoto\unskip,\iSLAC}
\mbox{D. Aston\unskip,\iSLAC}
\mbox{K.G. Baird\unskip,\iMASS}
\mbox{C. Baltay\unskip,\iYALE}
\mbox{H.R. Band\unskip,\iWISC}
\mbox{T.L. Barklow\unskip,\iSLAC}
\mbox{J.M. Bauer\unskip,\iMISSI}
\mbox{G. Bellodi\unskip,\iOXF}
\mbox{R. Berger\unskip,\iSLAC}
\mbox{G. Blaylock\unskip,\iMASS}
\mbox{J.R. Bogart\unskip,\iSLAC}
\mbox{G.R. Bower\unskip,\iSLAC}
\mbox{J.E. Brau\unskip,\iOREG}
\mbox{M. Breidenbach\unskip,\iSLAC}
\mbox{W.M. Bugg\unskip,\iTENN}
\mbox{D. Burke\unskip,\iSLAC}
\mbox{T.H. Burnett\unskip,\iWASH}
\mbox{P.N. Burrows\unskip,\iOXF}
\mbox{A. Calcaterra\unskip,\iFRAS}
\mbox{R. Cassell\unskip,\iSLAC}
\mbox{A. Chou\unskip,\iSLAC}
\mbox{H.O. Cohn\unskip,\iTENN}
\mbox{J.A. Coller\unskip,\iBU}
\mbox{M.R. Convery\unskip,\iSLAC}
\mbox{V. Cook\unskip,\iWASH}
\mbox{R.F. Cowan\unskip,\iMIT}
\mbox{G. Crawford\unskip,\iSLAC}
\mbox{C.J.S. Damerell\unskip,\iRAL}
\mbox{M. Daoudi\unskip,\iSLAC}
\mbox{S. Dasu\unskip,\iWISC}
\mbox{N. de Groot\unskip,\iBRI}
\mbox{R. de Sangro\unskip,\iFRAS}
\mbox{D.N. Dong\unskip,\iMIT}
\mbox{M. Doser\unskip,\iSLAC}
\mbox{R. Dubois\unskip,\iSLAC}
\mbox{I.Erofeeva\unskip,\iMOSCOW}
\mbox{V. Eschenburg\unskip,\iMISSI}
\mbox{E. Etzion\unskip,\iWISC}
\mbox{S. Fahey\unskip,\iCOLO}
\mbox{D. Falciai\unskip,\iFRAS}
\mbox{J.P. Fernandez\unskip,\iUCSC}
\mbox{K. Flood\unskip,\iMASS}
\mbox{R. Frey\unskip,\iOREG}
\mbox{E.L. Hart\unskip,\iTENN}
\mbox{K. Hasuko\unskip,\iTOHO}
\mbox{S.S. Hertzbach\unskip,\iMASS}
\mbox{M.E. Huffer\unskip,\iSLAC}
\mbox{X. Huynh\unskip,\iSLAC}
\mbox{M. Iwasaki\unskip,\iOREG}
\mbox{D.J. Jackson\unskip,\iRAL}
\mbox{P. Jacques\unskip,\iRUTG}
\mbox{J.A. Jaros\unskip,\iSLAC}
\mbox{Z.Y. Jiang\unskip,\iSLAC}
\mbox{A.S. Johnson\unskip,\iSLAC}
\mbox{J.R. Johnson\unskip,\iWISC}
\mbox{R. Kajikawa\unskip,\iNAGO}
\mbox{H.J. Kang\unskip, \iRUTG}
\mbox{M. Kalelkar\unskip,\iRUTG}
\mbox{R.R. Kofler\unskip,\iMASS}
\mbox{R.S. Kroeger\unskip,\iMISSI}
\mbox{M. Langston\unskip,\iOREG}
\mbox{D.W.G. Leith\unskip,\iSLAC}
\mbox{V. Lia\unskip,\iMIT}
\mbox{C.Lin\unskip,\iMASS}
\mbox{G. Mancinelli\unskip,\iRUTG}
\mbox{S. Manly\unskip,\iYALE}
\mbox{G. Mantovani\unskip,\iPERU}
\mbox{T.W. Markiewicz\unskip,\iSLAC}
\mbox{T. Maruyama\unskip,\iSLAC}
\mbox{A.K. McKemey\unskip,\iBRUN}
\mbox{R. Messner\unskip,\iSLAC}
\mbox{K.C. Moffeit\unskip,\iSLAC}
\mbox{T.B. Moore\unskip,\iYALE}
\mbox{M.Morii\unskip,\iSLAC}
\mbox{D. Muller\unskip,\iSLAC}
\mbox{V. Murzin\unskip,\iMOSCOW}
\mbox{S. Narita\unskip,\iTOHO}
\mbox{U. Nauenberg\unskip,\iCOLO}
\mbox{G. Nesom\unskip,\iOXF}
\mbox{N. Oishi\unskip,\iNAGO}
\mbox{D. Onoprienko\unskip,\iTENN}
\mbox{L.S. Osborne\unskip,\iMIT}
\mbox{R.S. Panvini\unskip,\iVAND}
\mbox{C.H. Park\unskip,\iSOONG}
\mbox{I. Peruzzi\unskip,\iFRAS}
\mbox{M. Piccolo\unskip,\iFRAS}
\mbox{L. Piemontese\unskip,\iFERR}
\mbox{R.J. Plano\unskip,\iRUTG}
\mbox{R. Prepost\unskip,\iWISC}
\mbox{C.Y. Prescott\unskip,\iSLAC}
\mbox{B.N. Ratcliff\unskip,\iSLAC}
\mbox{J. Reidy\unskip,\iMISSI}
\mbox{P.L. Reinertsen\unskip,\iUCSC}
\mbox{L.S. Rochester\unskip,\iSLAC}
\mbox{P.C. Rowson\unskip,\iSLAC}
\mbox{J.J. Russell\unskip,\iSLAC}
\mbox{O.H. Saxton\unskip,\iSLAC}
\mbox{T. Schalk\unskip,\iUCSC}
\mbox{B.A. Schumm\unskip,\iUCSC}
\mbox{J. Schwiening\unskip,\iSLAC}
\mbox{V.V. Serbo\unskip,\iSLAC}
\mbox{G. Shapiro\unskip,\iLBL}
\mbox{N.B. Sinev\unskip,\iOREG}
\mbox{J.A. Snyder\unskip,\iYALE}
\mbox{H. Staengle\unskip,\iCSU}
\mbox{A. Stahl\unskip,\iSLAC}
\mbox{P. Stamer\unskip,\iRUTG}
\mbox{H. Steiner\unskip,\iLBL}
\mbox{D. Su\unskip,\iSLAC}
\mbox{F. Suekane\unskip,\iTOHO}
\mbox{A. Sugiyama\unskip,\iNAGO}
\mbox{M. Swartz\unskip,\iJHU}
\mbox{F.E. Taylor\unskip,\iMIT}
\mbox{J. Thom\unskip,\iSLAC}
\mbox{T. Usher\unskip,\iSLAC}
\mbox{J. Va'vra\unskip,\iSLAC}
\mbox{R. Verdier\unskip,\iMIT}
\mbox{D.L. Wagner\unskip,\iCOLO}
\mbox{A.P. Waite\unskip,\iSLAC}
\mbox{S. Walston\unskip,\iOREG}
\mbox{J. Wang\unskip,\iSLAC}
\mbox{A.W. Weidemann\unskip,\iTENN}
\mbox{E. R. Weiss\unskip,\iWASH}
\mbox{J.S. Whitaker\unskip,\iBU}
\mbox{S.H. Williams\unskip,\iSLAC}
\mbox{S. Willocq\unskip,\iMASS}
\mbox{R.J. Wilson\unskip,\iCSU}
\mbox{W.J. Wisniewski\unskip,\iSLAC}
\mbox{J. L. Wittlin\unskip,\iMASS}
\mbox{M. Woods\unskip,\iSLAC}
\mbox{T.R. Wright\unskip,\iWISC}
\mbox{R.K. Yamamoto\unskip,\iMIT}
\mbox{J. Yashima\unskip,\iTOHO}
\mbox{S.J. Yellin\unskip,\iUCSB}
\mbox{C.C. Young\unskip,\iSLAC}
\mbox{H. Yuta\unskip.\iAOMORI}

\it
\vskip \baselineskip                   
\vskip \baselineskip      
\baselineskip=.75\baselineskip   
\iAOMORI
Aomori University, Aomori , 030 Japan, \break
\iBRI
University of Bristol, Bristol, United Kingdom, \break
\iBRUN
Brunel University, Uxbridge, Middlesex, UB8 3PH United Kingdom, \break
\iBU
Boston University, Boston, Massachusetts 02215, \break
\iCOLO
University of Colorado, Boulder, Colorado 80309, \break
\iCSU
Colorado State University, Ft. Collins, Colorado 80523, \break
\iFERR
INFN Sezione di Ferrara and Universita di Ferrara, I-44100 Ferrara, Italy, \break
\iFRAS
INFN Lab. Nazionali di Frascati, I-00044 Frascati, Italy, \break
\iJHU
Johns Hopkins University,  Baltimore, Maryland 21218-2686, \break
\iLBL
Lawrence Berkeley Laboratory, University of California, Berkeley, California 94720, \break
\iMASS
University of Massachusetts, Amherst, Massachusetts 01003, \break
\iMISSI
University of Mississippi, University, Mississippi 38677, \break
\iMIT
Massachusetts Institute of Technology, Cambridge, Massachusetts 02139, \break
\iMOSCOW
Institute of Nuclear Physics, Moscow State University, 119899, Moscow Russia, \break
\iNAGO
Nagoya University, Chikusa-ku, Nagoya, 464 Japan, \break
\iOREG
University of Oregon, Eugene, Oregon 97403, \break
\iOXF
Oxford University, Oxford, OX1 3RH, United Kingdom, \break
\iPERU
INFN Sezione di Perugia and Universita di Perugia, I-06100 Perugia, Italy, \break
\iRAL
Rutherford Appleton Laboratory, Chilton, Didcot, Oxon OX11 0QX United Kingdom, \break
\iRUTG
Rutgers University, Piscataway, New Jersey 08855, \break
\iSLAC
Stanford Linear Accelerator Center, Stanford University, Stanford, California 94309, \break
\iSOONG
Soongsil University, Seoul, Korea 156-743, \break
\iTENN
University of Tennessee, Knoxville, Tennessee 37996, \break
\iTOHO
Tohoku University, Sendai 980, Japan, \break
\iUCSB
University of California at Santa Barbara, Santa Barbara, California 93106, \break
\iUCSC
University of California at Santa Cruz, Santa Cruz, California 95064, \break
\iVAND
Vanderbilt University, Nashville,Tennessee 37235, \break
\iWASH
University of Washington, Seattle, Washington 98105, \break
\iWISC
University of Wisconsin, Madison, Wisconsin 53706, \break
\iYALE
Yale University, New Haven, Connecticut 06511. \break

\rm
%

\end{center}

The SLD Collaboration has performed a series of 
increasingly precise measurements
of the left-right cross-section asymmetry in the
production of $\zo$ bosons by \ee\ collisions \cite{alr92,alr93,alr95}.
In this letter,
we present a measurement based upon data
recorded during the 1996 and 1997-98 runs of the 
SLAC Linear Collider (SLC), which represents about
three quarters of our total sample and leads to
improved statistical 
precision and reduced systematic uncertainty.
The overall average given at the end of this Letter
is based upon all the data from the
completed SLD experimental program \cite{PRD}.

The left-right asymmetry is defined as
$\alro\equiv\left(\sigma_L-\sigma_R\right)/
\left(\sigma_L+\sigma_R\right)$,
where $\sigma_L$ and $\sigma_R$ are the $\ee$ production
cross sections for $\zo$ bosons at the $\zo$-pole energy
with left-handed and right-handed
electrons, respectively.  The Standard Model predicts
that this quantity depends upon the effective vector ($v_e$)
and axial-vector ($a_e$) couplings of the $\zo$ boson to the electron
current,
\begin{equation}
\alro=\frac{2v_ea_e}{v_e^2+a_e^2}\equiv \frac{
2\left[1-4\swein\right]}{1+\left[1-4\swein\right]^2}, \label{eq:alrswein}
\end{equation}
where the effective electroweak mixing parameter
is defined \cite{swdef}
as $\swein\equiv(1-v_e/a_e)/4$.
The quantity $\alro$ is a sensitive function of $\swein$
and depends upon virtual
electroweak radiative corrections including those which involve
the Higgs boson and those arising from new phenomena
outside of the scope of the Standard Model (SM).
Presently, the most stringent upper bounds on the SM Higgs mass
are provided by measurements of $\swein$.

We measured the left-right asymmetry by counting hadronic and (with low efficiency) $\tau^+\tau^-$ final states produced in \ee\ collisions
near the $\zo$-pole energy for each of the two longitudinal polarization
states of the electron beam.  The asymmetry formed from these rates,
$\alr$, was then corrected for residual effects arising from 
pure photon
exchange and $\zo$-photon interference 
to extract $\alro$.  The measurement required
knowledge of the absolute beam polarization, but
did not require knowledge of the
absolute luminosity,
detector acceptance, or efficiency \cite{accept}.

The operation of the SLC with a polarized electron beam
has been described previously \cite{oldslc}. 
The maximum luminosity of the collider was approximately
3$\times$10$^{30}$~cm$^{-2}$sec$^{-1}$, and the longitudinal electron
polarization at the $\ee$ collision point was typically $\sim$75\%. 
The luminosity-weighted mean $\ee$ center-of-mass energy ($E_{cm}$) was
measured with
precision energy spectrometers \cite{enspa} 
and was found to be 91.26$\pm$0.03~GeV for the 1996 run. 
During the 1997-98 period,  
the energy spectrometers were (for the first time)
calibrated to the well-measured $\zo$ boson mass \cite{lepew}
by performing a three-point scan of the
resonance \cite{ZFITTERFIT}, 
with the result $E_{cm} = 91.237\pm0.029$~GeV for the 1997-98 run.

The longitudinal electron beam polarization ($\pole$)
was measured by a Compton scattering
polarimeter \cite{polarimeter},\cite{alr92,alr93,alr95}. 
The primary device was a magnetic spectrometer and multichannel Cherenkov
detector that observed Compton-scattered electrons in the energy 
range 17 GeV to 30 GeV.
The analyzing powers of the detector 
channels incorporated resolution and spectrometer effects, 
and differed by typically $\sim 1\%$
from the theoretical Compton polarization
asymmetry function \cite{comref} at the mean accepted energy
for each channel.
The minimum energy of a Compton-scattered electron 
for the initial electron and photon energies was 17.36~GeV.  
The location of this kinematic endpoint at the detector (in the dispersive plane of the spectrometer) was monitored by
frequent scans of the detector horizontal position during polarimeter 
operation.  This technique determined and monitored the analyzing powers of 
each detector channel.
Polarimeter data were acquired continually during the operation
of the SLC.  

Beginning in 1996, two additional detectors were operated in order to
assist in the calibration of the primary spectrometer-based polarimeter.
Both devices detected Compton-scattered photons 
and hence were independent of the spectrometer 
calibration and its systematic uncertainties. 
Due to their inherent
sensitivity to beamstrahlung background,
these two devices, the Polarized Gamma Counter(PGC) \cite{PGC}
and the Quartz Fiber Calorimeter(QFC) \cite{QFC},
were operated only when the
electron and positron beams were not in collision.  
However, when compared with
concurrent results from the primary detector they 
achieved comparable precision and provided 
a useful crosscheck of our calibration procedure.

The systematic uncertainties that affect the polarization measurement
are summarized in Table~\ref{table1}.  The largest contribution, due
to analyzing power calibration, was estimated by a comparison of
our reference polarization measurement provided by the Cherenkov
detector channel located at the kinematic endpoint (and Compton
asymmetry maximum) to the results from a neighboring channel and from
the PGC and QFC devices.  A $\sim 0.6\%$ systematic error on the PGC 
calibration was dominated by the difference in the photon energy response 
function as determined from test beam data, and from EGS \cite{EGS} 
Monte Carlo simulations. 
For the QFC device, 
uncertainties on the linearity of the response function, also deduced from 
test beam data, dominated the total systematic error of $\sim 0.6\%$. 
The weighted mean residual of all analyzing power cross checks is 
$0.30\% \pm 0.39\%$ ($\chi^2=1.9$ for 2 degrees of freedom), from which 
we quote a calibration uncertainty of $0.4\%$.

Interspersed high- and low-background polarimeter operation in 1997-98, 
achieved by periodic removal of the positron beam,
permitted improved
studies of the Cherenkov detector linearity and
significantly reduced the associated uncertainty, previously our
largest effect, to 0.2\% \cite{colldepol}.  
The total relative systematic uncertainty is 
estimated to be $\delta\pole/\pole=0.50\%$ (down from $0.65\%$ \cite{alr95}).

In our previous Letters \cite{alr93,alr95},
we examined an effect that causes the beam polarization
measured by the Compton Polarimeter, $\pole$, to
differ from the luminosity-weighted beam polarization,
$\pole(1+\xi)$, at the SLC interaction point (IP), where
$\xi$ is a small fractional correction. 
A number of measures in the operation of the SLC 
and in monitoring procedures  
reduced the size of this {\it chromaticity}
correction and its associated error to 
below $0.2\%$ \cite{chromat}.
From beam energy spread, polarization transport, and
luminosity energy dependence measurements,
we determined a contribution to $\xi$ of
$+0.00124\pm0.0012$ (1996) and
$+0.00117\pm0.0008$ (1997-98)
due to the chromaticity effect.  The results for both runs are smaller
than for previous years \cite{alr95}.

A similar effect of comparable magnitude
arises due to the small
precession of the electron spin in the final focusing
elements between the SLC IP and the polarimeter.
We estimated  
this effect contributed $-0.0011\pm0.0005$ to $\xi$ in 1996, and 
$-0.0024\pm0.0008$ to $\xi$ in 1997-98, where the larger
value in the recent data reflects the larger focusing angles
used at the time. 

The depolarization of the electron beam by the \ee\ collision 
process is expected to be negligible \cite{chenyok}.  
The contribution of depolarization to $\xi$ was determined 
to be 0.000$\pm$0.001 by comparing polarimeter data
taken with and without beams in collision.
Combining the three effects described above, the overall correction
factors were determined to be 
$\xi = 0.0002\pm0.0016$ (1996) and
$\xi = -0.0012\pm0.0015$ (1997-98).

The $\ee$ collisions were measured by the SLD detector
which has been described elsewhere \cite{sld}.
For $\zo$ decays
the detector trigger and the event selection relied on the
liquid argon calorimeter (LAC) \cite{lac} and the central drift
chamber tracker (CDC) \cite{cdc}.
For each event candidate,
energy clusters were reconstructed in the LAC.  Selected
events were required to contain at least 22~GeV of energy observed in the
clusters and to manifest a normalized energy
imbalance of less than 0.6 \cite{eimb}.  The left-right asymmetry associated
with final state $\ee$ events is expected to be diluted by the t-channel
photon exchange subprocess.  Therefore, we excluded $\ee$ final states
by requiring that each event candidate contain at least 4
selected CDC tracks,
with at least 2 tracks in each hemisphere (defined with respect to
the beam axis), or at least 4 tracks in either hemisphere.
This track topology requirement excludes Bhabha events which contain
a reconstructed gamma conversion.
The selected CDC tracks were required to extrapolate to the IP within 
5 (10) cm radially (along the beam direction), to have
a minimum momentum transverse to the beam direction of 100 MeV/c,
and to form a minimum angle of 30 degrees with the beam direction.

We estimate that the combined efficiency of the trigger
and selection criteria was (91$\pm$1)\% for
hadronic $\zo$ decays.  Tau pairs constituted (0.3$\pm$0.1)\% of the sample.
Because muon pair events deposited little energy in the calorimeter,
they were not included in the sample.
A residual background
in the sample was due to
$\ee$ final state events.
We use our data and a Monte Carlo simulation to
estimate this background fraction
to be ($0.013\pm 0.013)\%$.  The background fraction
due to cosmic rays, two-photon events and beam related processes
was estimated to be (0.029$\pm$0.029)\% for 1997-98, and 
(0.016$\pm$0.016)\% for 1996.

For the 1997-98 (1996) datasets respectively, 
a total of 331,614 (51,873) $\zo$ events
satisfied the selection criteria.
We found that 183,355 (29,016) of the events
were produced with the left-handed
electron beam ($N_L$) and 148,259 (22,857)
were produced with the right-handed beam ($N_R$).
The measured left-right cross-section asymmetry
is \cite{helicity}
\begin{eqnarray*}
\ameas\equiv\frac{N_L-N_R}{N_L+N_R}=\left \{ \begin{array} {ll} 0.10583\pm0.00173 & 97/8 \\ 0.11873\pm0.00436 &
96. \end{array} \right . 
\end{eqnarray*}
We verified that the measured asymmetry $\ameas$
did not vary significantly as more restrictive criteria (calorimetric
and tracking-based) were applied to the sample and that $\ameas$ was
uniform when binned by the azimuth and polar angle of the thrust axis.

The measured asymmetry $\ameas$ is related to $\alr$ by the following
expression which incorporates a number of small correction terms in
lowest-order approximation,
\begin{eqnarray}
\alr & = & \frac{\ameas}{\apolel}+\frac{1}
{\apolel}\biggl[f_b(\ameas-\aback)-\alum+\ameas^2\apol \nonumber \\
&  & -E_{cm}\frac{\sigma^\prime(E_{cm})}{\sigma(E_{cm})}\aengy
-\aeff + \apolel\polp \biggr], \label{eq:alrcor}
\end{eqnarray}
where $\apolel$ is the mean luminosity-weighted polarization; 
$f_b$ is the background fraction;
$\sigma(E)$ is the unpolarized $\zo$ cross section at energy $E$;
$\sigma^\prime(E)$ is the derivative of the cross section with
respect to $E$;
$\aback$, $\alum$, $\apol$, $\aengy$, and
$\aeff$ are the left-right asymmetries \cite{asymdef}
of the residual background,
the integrated luminosity, the beam polarization,
the center-of-mass energy, and
the product of detector acceptance and efficiency, respectively;
and $\polp$ is any longitudinal positron polarization which is assumed to
have constant helicity \cite{ppol}.

In the past, we have taken $\polp$ to be negligible, based on
calculations of transverse polarization buildup in the SLC
positron damping ring (ignoring
efficiencies in positron polarization transport to the
beam collision point) 
that indicate the effect cannot be larger
than a few parts in $10^5$.
Nevertheless, we determined that we could address this issue
experimentally, and directly measured
$\polp$ in 1998.  The SLC positron beam was delivered to the
fixed target M\o ller polarimeter in SLAC's End Station A
\cite{ESApol} in a one week dedicated experiment, and the
result ($\polp$ = -0.02$\pm$0.07\%) was consistent
with zero \cite{posipol}.

The luminosity-weighted
average polarization $\apolel$ 
for the 1997-98 (1996) data
was estimated from measurements of
$\pole$ made when $\zo$ events were recorded,
\begin{equation}
\apolel = (1+\xi)\cdot\frac{1}{N_Z}\sum_{i=1}^{N_Z}{\poll}_i
=\left \{ \begin{array} {ll} 72.92\pm0.38\% & 97/8 \\ 76.16\pm0.40\% &
96, \end{array} \right . \label{eq:poldef}
\end{equation}
where
$N_Z$ is the total number of $\zo$ events, and ${\poll}_i$ is the
polarization measurement associated in time with the $i^{th}$ event.
The error on $\apolel$ was dominated by the systematic
uncertainties on the polarization measurement.  The 
different values for $\apolel$ seen during different
SLC running periods are due to different
GaAs photocathodes used at the SLC polarized source.

The corrections defined in equation~(\ref{eq:alrcor}) were found to be small.
The results for 1997-98 (1996) are detailed below.
The correction for residual background contamination was moderated by
a non-zero left-right background asymmetry [$A_b=0.023\pm0.022$
(0.033$\pm$0.026)]
arising from $\ee$ final states which remained in the sample.
Residual electron current asymmetry ($\lsim10^{-3}$) from the SLC polarized source was reduced by periodically reversing
a spin rotation solenoid at the entrance to the SLC
damping ring.  The net luminosity asymmetry was 
estimated from the measured asymmetry of the rate of radiative Bhabha
scattering events observed with a monitor located in the North
Final Focus region of the SLC
to be $\alum=[-1.3\pm0.7]\times10^{-4} ([+0.03\pm0.5]\times10^{-4})$.
A statistically less precise cross check was performed
by examining the left-right asymmetry of the sample of approximately
800,000 small-angle
Bhabha scattering events detected by
the luminosity monitoring system (LUM) \cite{berridge}.
Since the theoretical left-right asymmetry for small-angle
Bhabha scattering is very small [${\cal O} (10^{-4})\pole$
within the LUM acceptance], the measured
asymmetry of [$-$10$\pm$10]$\times$10$^{-4}$ was a direct determination
of $\alum$ and was consistent with the more precisely determined one.
The polarization asymmetry was directly measured to be
$\apol=[+2.8\pm6.9]\times10^{-3} ([+2.9\pm4.3]\times10^{-3})$.
The left-right beam energy asymmetry arises
from the small residual left-right
beam current asymmetry due to beam-loading of the accelerator and was
measured to be [+2.8$\pm$1.4]$\times$10$^{-7}$ 
([-0.1$\pm$3.5]$\times$10$^{-7})$.  
The coefficient of the energy
asymmetry in equation~(\ref{eq:alrcor}) 
is a very sensitive function of the 
center-of-mass energy and was found to be $4.3\pm2.9$ for
$E_{cm}=91.237\pm0.029$~GeV ($2.0\pm3.0$ for
$E_{cm}=91.26\pm0.03$~GeV).
The SLD had a symmetric acceptance in polar angle \cite{accept} which implied
that the efficiency asymmetry $\aeff$ is negligible.
The corrections listed in equation~(\ref{eq:alrcor}) change
$\alr$ by [$+0.16\pm 0.07$]\% ([$+0.02\pm 0.05$]\%) of the uncorrected value.

From equation~(\ref{eq:alrcor}), we found the left-right asymmetry
to be $\alr(91.237~{\rm GeV}) =
0.1454\pm0.00237({\rm stat.})\pm0.00077({\rm syst.})$, for 1997-98 and
$\alr(91.26~{\rm GeV}) =
0.1559\pm0.00572({\rm stat.})\pm0.00084({\rm syst.})$
for 1996.

We found the pole asymmetry $\alro$ for 1997-98 to be
$\alro = 0.14906\pm0.00237({\rm stat.})\pm0.00096({\rm syst.})$,
and
$\alro = 0.15929\pm0.00573({\rm stat.})\pm0.00101({\rm syst.})$,
for 1996, where the systematic uncertainty includes the 
uncertainty on the electroweak interference correction 
(see Table~\ref{table1}) which arose from the uncertainty 
on center-of-mass energy scale.
Combining the value of $\alro$ and
$\swein$ \cite{ewcorr} provided by the 1996-98
data of $\alro=0.15056\pm0.00239$ and $\swein=0.23107\pm0.00030$
with our previous measurements
 \cite{alr92,alr93,alr95} (systematic errors are conservatively
taken to be fully correlated between measurements) we obtain the value,
$$ \alro = 0.15138\pm0.00216 \qquad \swein = 0.23097\pm0.00027. $$
This $\swein$ determination is the most precise presently
available, and is smaller by
2.7 standard deviations than the recent average of measurements performed
by the LEP Collaborations \cite{lepew}.

We thank the personnel of the SLAC accelerator department
and the technical staffs of our collaborating institutions for their
outstanding efforts on our behalf.
This work was supported by the Department of Energy; the
National Science Foundation; the Istituto Nazionale di Fisica
Nucleare of Italy;
the Japan-US Cooperative Research Project on High Energy Physics;
and the Science and Engineering Research Council of the United Kingdom.


%
%


%
%
\begin{table}
\caption{Systematic uncertainties that affect the $\alr$ measurement.  The uncertainty on the electroweak interference correction
is caused by the uncertainty on the SLC energy scale.  Where they differ from the errors for the
1997/98 data, the errors for 1996 are given in parentheses. }
\label{table1}
\begin{tabular}{lccc}
Uncertainty & $\delta\pole/\pole$~(\%) & $\delta\alr/\alr$~(\%) & $\delta\alro/\alro$~(\%)\\
\hline
Laser Pol. & 0.10 & & \\
Linearity & 0.20 & & \\
Anal. Pwr. Cal. & 0.40 & & \\
Electr. Noise & 0.20 & & \\ \hline
Total Polarim.  & 0.50 & 0.50 & \\
$\xi$ (Eq.~\ref{eq:poldef}) &     & 0.15(0.16) & \\
Corrs in Eq.~\ref{eq:alrcor} &  & 0.07(0.05)& \\ \hline
$\alr$ Total &    & 0.52(0.52) & 0.52(0.52)\\
EW Int. Corr. &    &  & 0.39(0.37)\\ \hline
$\alro$ Total  &    &  & 0.64(0.63)\\
\end{tabular}
\end{table}


\begin{references}
\bibitem{accept} The value of $\alr$ is unaffected by decay-mode-dependent
variations in detector acceptance and efficiency
provided that the efficiency for detecting a fermion at
some polar angle (with respect to the electron direction)
is equal to the efficiency for detecting
an antifermion at the same polar angle.
\bibitem{colldepol} The assumption that collisional depolarization
was negligible was implicit
in our comparison of high and low background running used to
constrain nonlinearities in our polarimeter response. Our  
experimental test of this assumption is
described later in this Letter.
\bibitem{eimb} The energy imbalance is defined as a normalized vector
sum of the energy clusters as follows,
$ E_{imb}=|\sum \vec E_{cluster}|/\sum |E_{cluster}|$.
\bibitem{helicity} The absolute sign of $\ameas$ is inferred from the sign of the
measured Compton scattering asymmetry, the
measured helicity of the polarimeter laser, and the theoretical sign
of the Compton scattering asymmetry.
\bibitem{asymdef} The left-right asymmetry for a quantity $Q$ is defined as
$A_Q\equiv(Q_L-Q_R)/(Q_L+Q_R)$ where the subscripts $L$,$R$ refer to
the left- and right-handed beams, respectively.
\bibitem{ppol} Since the colliding electron and positron bunches
were produced on different machine cycles and since the
electron helicity of each cycle was chosen randomly, any positron
helicity arising from the polarization of the production electrons
was uncorrelated with electron helicity at the IP.  The net positron
polarization from this process vanished rigorously.  However,
positron polarization of constant helicity would affect the measurement.
\bibitem{ewcorr} The quantities $\alro$ and $\swein$ are related by
equation~(\ref{eq:alrswein}) and are completely equivalent.  
The correction for pure photon exchange and
for electroweak interference (which arises from
the deviation of the effective \ee\ center-of-mass energy
from the $\zo$-pole energy), including the effect of initial-state
radiation, 
$\alro-\alr(E_{cm})$, is determined with the ZFITTER~6.22
program (see reference~\cite{ZFITTERFIT})
and is found to be $0.00358\pm0.00058$ ($0.00337\pm0.00059$).
\end{references}
\end{document}